\documentstyle[epsfig,pra,aps,tighten,twocolumn]{revtex}
\newcommand{\be}{\begin{equation}}
\newcommand{\ee}{\end{equation}}
\newcommand{\bea}{\begin{eqnarray}}
\newcommand{\eea}{\end{eqnarray}}
\newcommand{\ba}{\begin{array}}
\newcommand{\ea}{\end{array}}
\newcommand{\nn}{\nonumber}
\begin{document}
\draft
\title{\bf Superluminal Optical Phase Conjugation: Pulse Reshaping and 
Instability}

\author{M. Blaauboer,$^{\rm a}$  A.G. Kofman, $^{\rm b}$  A.E. Kozhekin, 
$^{\rm b}$
G. Kurizki, $^{\rm b}$ D. Lenstra, $^{\rm a}$  and  A. Lodder$^{\rm a}$} 

\address{$^a$Faculteit Natuurkunde en Sterrenkunde, Vrije Universiteit,
         De Boelelaan 1081, 1081 HV Amsterdam, The Netherlands \\
$^b$Chemical Physics Department, Weizmann Institute of Science, 
Rehovot 76100, Israel}

\date{\today}
\maketitle

\begin{abstract}
We theoretically investigate the response of optical phase conjugators
to incident probe pulses. In the stable (sub-threshold) operating 
regime of an optical phase conjugator it is possible to transmit 
probe pulses with a superluminally advanced peak, whereas 
conjugate reflection is always subluminal. In the unstable 
(above-threshold) regime, superluminal 
response occurs both in reflection and in transmission, at times preceding 
the onset of exponential growth due to the instability. 
\end{abstract}

\pacs{PACS numbers: 42.65.Hw, 42.25.Bs, 42.50.Md
                {\tt physics/9803015}}
\narrowtext

\section{Introduction}

A variety of mechanisms is now known to give rise to superluminal
(faster-than-$c$) group velocities, which express the peak advancement 
of electromagnetic pulses reshaped by material media:
\begin{enumerate}
\item{{\it Near-resonant absorption}\cite{brillouin60,chu82}: 
anomalous dispersion in the linear regime of an absorbing medium 
forms the basis for this superluminal reshaping mechanism.}
\item{{\it Reduced transmission or evanescent wave formation (tunneling) in
passive dielectric structures}\cite{buettiker82,steinberg92,ranfagni93}:
this reshaping mechanism has been attributed to interference between 
multiply-reflected 
propagating pulse components in the structure\cite{japha96}.} 
\item{{\it Soliton propagation in dissipative nonlinear structures} 
\cite{picholle91}: 
superluminal group velocities can occur in such systems via nonlinear 
three-wave exchanges, as in stimulated Brillouin backscattering in the
presence of dissipation. They also occur in a nonlinear
laser amplifier\cite{icsevgi69}.}
\item{{\it Pulse propagation in transparent (non-resonant) amplifying
media}\cite{bolda96,artoni98}. Superluminal pulse reshaping
in this regime has been attributed to either the dispersion\cite{bolda96}
or the boundary reflections\cite{artoni98} of the amplifying medium.}
\item{{\it Tachyonic dispersion in inverted two-level media}:
the dispersion in such inverted {\it collective} systems is analogous to the 
tachyonic dispersion exhibited by a Klein-Gordon particle with 
imaginary mass\cite{aharonov69}. Consequently, it has been suggested
\cite{chiao96} that probe pulses in such media can exhibit 
superluminal group velocities provided they are spectrally narrow. 
Gain and loss have been assumed to be detrimental for such reshaping.
We note that Ref.~\cite{chiao96} describes an infinite medium 
and boundary effects on the reshaping have not been considered.} 
\end{enumerate}

In this paper we establish a connection between optical
phase conjugation and tachyonic-type behavior in a finite medium.
To that end we study the reflection and transmission of pulses at 
a phase-conjugating
mirror (PCM), both in its stable and in its unstable
operating regimes. Similar features in other 
parametric processes, such as stimulated Raman scattering and parametric
down-conversion will be treated elsewhere. The dispersion
relation in {\it infinite} PCM media, which allows for superluminal 
group velocities, was derived by Lenstra\cite{lenstra90}. 
Here we address the questions: can superluminal 
features be observed in the response of a {\it finite} PCM 
to an incident probe pulse, and how can they be reconciled with causality?

Pulse reshaping by a PCM has been studied before\cite{fisher81}, but 
never in the context of superluminal behavior. We show 
(Sec.~\ref{sec:tempB}) that 
in the well-known stable operating regime of a PCM, superluminal peak 
advancement can occur in transmission but not in conjugate reflection, 
as was briefly presented in Ref.~\cite{blaauboer97}.
In the unstable regime (Sec.~\ref{sec:tempC}), we demonstrate 
the existence of superluminal peak advancement in both the reflected 
and transmitted wave response to spectrally narrow analytic probes.
Further insight into these processes is gained by spatial
wavepacket analysis (Sec.~\ref{sec:movie}).
For sharply-timed (abruptly modulated, non-analytic) 
signals (Sec.~\ref{sec:chop}) the occurrence of spectral 
components in the frequency zone where the gain 
is highest cannot be avoided. These components trigger 
early exponential growth, which is independent of the probe pulse shape. 
In all cases we develop criteria for the optimization of probe pulse shapes, 
aimed at maximizing their superluminal peak advancement and, in the 
unstable regime, the delay of the instability onset.

\begin{figure}
\centerline{\epsfig{figure=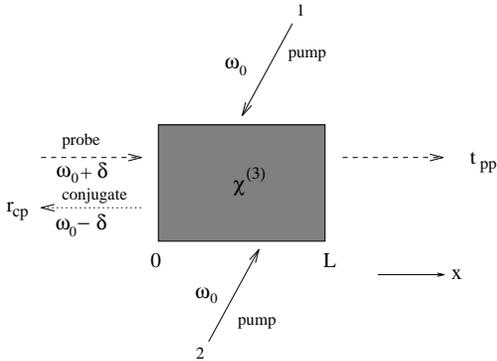,width=0.75\hsize}}
\caption{Phase conjugation by four-wave mixing.
Solid (dashed, dotted) arrows denote pump (probe, conjugate) beams. 
}
\label{fig:fwm}
\end{figure}

Our phase-conjugating mirror consists of a nonlinear optical medium with 
a large third-order susceptibility $\chi^{(3)}$, confined to a cell of length
$L$. Phase conjugation is obtained through a four-wave mixing (FWM) process
\cite{fisher83}: when the medium is pumped by two intense counterpropagating 
laser beams of frequency $\omega_{0}$ and a weak probe beam of frequency 
$\omega_{p} = \omega_{0} + \delta$ is incident on the cell, a fourth beam 
will be generated 
due to the nonlinear polarization of the medium. This conjugate wave
propagates with frequency $\omega_{c} = \omega_{0} - \delta$ in 
the opposite 
direction as the probe beam (see Fig.~\ref{fig:fwm}).

\section{Pulse reflection and transmission - temporal analysis}
\label{sec:temp}

\subsection{Basic Analysis}
\label{sec:tempA}

The basic semiclassical one-dimensional equations describing this FWM process 
are obtained by substituting the total field $E(x,t) = 
\sum_{\alpha=1,2,p,c} E_{\alpha} (x,t) = \sum_{\alpha=1,2,p,c}
{\cal E}_{\alpha}(x,t) e^{i(k_{\alpha}x - \omega_{\alpha} t)} + 
c.c.$ (where the labels
1,2 refer to the two pump beams and p,c to the probe and conjugate
beams respectively) into the wave equation for a nonmagnetic, 
nondispersive material in the presence of a nonlinear polarization
\be
\left( \frac{\partial^2}{\partial x^2} - \epsilon_{r} \epsilon_{0} \mu_{0}
\frac{\partial^2}{\partial t^2} \right) E(x,t) = \frac{1}{\epsilon_{0} c^2}
\frac{\partial^2}{\partial t^2} P_{NL}(x,t).
\label{eq:waveeqn}
\ee
Selecting the phase-conjugation terms for $P_{NL}(x,t) = \chi^{(3)} E^3(x,t)$,
assuming the pump beams to 
be non-depleted and applying the slowly varying envelope approximation
(SVEA) results in\cite{fisher83}
\be
\left( \begin{array}{cc}
\frac{\partial}{\partial x} +
\frac{1}{c} \frac{\partial}{\partial t}  & i\, \kappa \vspace{0.3cm} \\
i\, \kappa^{*} &  \frac{\partial}{\partial x} 
- \frac{1}{c} \frac{\partial}{\partial t} \end{array} \right)
\left( \begin{array}{l}
{\cal E}_{p}(x,t) \vspace{0.3cm} \\ {\cal E}_{c}^{*}(x,t)
\end{array} \right) = 0.
\label{eq:SEFL}
\ee
Here ${\cal E}_{p(c)}^{(*)}(x,t)$ denotes the complex amplitude of the probe
(conjugate) field and $\kappa \equiv \kappa_{0} e^{i\phi} = \frac{3\omega_{0}}
{\epsilon_{0}c} \chi^{(3)} {\cal E}_{1} {\cal E}_{2}$ is the 
coupling strength (per unit length) between the probe and conjugate wave.
The dispersion relation for an electromagnetic
excitation in this pumped nonlinear medium is given by
\cite{lenstra90} 
\be
k^2 = \frac{\omega_{0}^2}{c^2} \pm \frac{2\omega_{0}}{c^2} 
\sqrt{\delta^2 + \kappa_{0}^2 c^2}, 
\label{eq:tachdisp}
\ee
where $\delta = \omega_{p} - \omega_{0} = \omega_{0} - \omega_{c}$, and
$k^2$ is the squared wavevector of both the probe and the conjugate waves.

The group velocity $\partial \delta/ \partial k$ in the medium can be 
shown from (\ref{eq:tachdisp}) to be always
larger than {\it c}, the speed of light in vacuum, and the dispersion relation
is therefore of the tachyonic type (see Fig.~\ref{fig:gaussstable}(a)),
analogous to the one for inverted 
atoms\cite{chiao96}. Because of the superluminal group velocity caused
by this dispersion, the question
arises how wave packets will be reshaped by a PCM.
In the early 80's, Fisher {\it et al.} studied the phase-conjugate 
reflection of pulses of arbitrary shape at a PCM\cite{fisher81}.
No tachyonic effects were found.
They distinguished between two different operating regimes of the mirror, 
stable (if $\kappa_{0}L < \frac{\pi}{2}$) and unstable (if 
$\kappa_{0}L > \frac{\pi}{2}$). In the stable regime, the response 
of the mirror is always finite and scales with the probe input. 
The unstable regime corresponds to self-generation of conjugate reflection
from arbitrarily small probe input, followed by exponential growth
until saturation is reached (due to depletion).
Recently we predicted the occurrence of
superluminal advancement of the peak of a suitably chosen input pulse
upon transmission through a stable PCM\cite{blaauboer97}. 
Here we present the full analysis of both the stable and the
unstable regimes and demonstrate the existence of superluminal
effects in both.

In order to study pulse reflection and transmission at a PCM, we use
the two-sided Laplace transform (TSLT) technique introduced by Fisher 
{\it et al.}\cite{fisher81}. The basic approach is summarized in App.~A.
To begin with, the reflection and transmission amplitudes for a
monochromatic probe beam incident on a PCM 
are given by (Eqs.~(\ref{eq:conjrefl3}) and (\ref{eq:probetrans3}),
at $x=0$ and $x=L$, respectively)
\begin{eqnarray}
r_{cp} (\delta) & = & \frac{\kappa_{0} \sin(\beta L)} {\frac{\delta}{c}
\mbox{\rm sin} (\beta L) + i \beta \mbox{\rm cos} (\beta L)}
\label{eq:conjrefl} \\
t_{pp} (\delta) & =  & \frac{i \beta}
{\frac{\delta}{c}\mbox{\rm sin} (\beta L) + i \beta \mbox{\rm cos} 
(\beta L)} 
\label{eq:probetrans}
\end{eqnarray}
where
\be
\beta = \frac{1}{c} \sqrt{\delta^2 + (\kappa_{0}\, c)^2}.
\label{eq:betavec}
\ee
Now consider a probe field ${\cal E}_{p}(0,t)$ 
incident on the PCM at $x=0$, which satisfies the basic TSLT 
premise that it decreases faster than exponentially as 
$t\to -\infty$\cite{laplbook}.
We obtain the expressions for the resulting reflected phase-conjugate 
pulse at $x=0$, and the
transmitted probe pulse at $x=L$ from the inverse TSLT (see App.~A). In taking 
the inverse TSLT we separate the singularities of $r_{cp}(is)$ and
$t_{pp}(is)$ in the right-half $s$-plane, which give rise to unstable 
exponentially growing solutions, from the singularities in the left-half
$s$-plane, which correspond to stable solutions. The final expressions
for $ 0 <\! \kappa_{0}\, \! L < \frac{3\pi}{2}$ are
\begin{eqnarray}
{\cal E}_{c}(0,t) & = & \frac{1}{2\pi i} \int_{- i\infty}^{i\infty} 
ds\, \bar{r}_{cp} (i s)\, \tilde{{\cal E}}_{p^{*}}(0,s)\, e^{st}  \nonumber \\
&& + \  h_{1}\, \int_{- \infty}^{t}\, dt^{'}\, {\cal E}_{p}^{*}(0,t^{'})\,
e^{-s_{1}(t^{'} - t)}, 
\label{eq:fisherr2}
\end{eqnarray}
\begin{eqnarray}
{\cal E}_{p}(L,t) & = & \frac{1}{2\pi i} \int_{- i\infty}^{i\infty} 
ds\, \bar{t}_{pp} (i s)\, \tilde{{\cal E}}_{p}(0,s)\, e^{st} \nonumber \\
&& + \ h_{2}\, \int_{- \infty}^{t}\, dt^{'}\, {\cal E}_{p}(0,t^{'})\,
e^{-s_{1}(t^{'} - t)}.
\label{eq:fishert2}
\end{eqnarray}
Here
\begin{eqnarray}
\bar{r}_{cp} (is) & \equiv & r_{cp}(is) - \frac{h_{1}}{s - s_{1}} \\
\bar{t}_{pp} (is) & \equiv & t_{pp}(is) - \frac{h_{2}}{s - s_{1}} \\
s_{1} & = & \left| c \, \kappa_{0}\, \mbox{\rm cos} X_{1} \right| \\
h_{1} & = & -i\, c\, \kappa_{0} \, \mbox{\rm sin}^2 (X_{1})/
[1 + \frac{L s_{1}}{c}]  \\
h_{2} & = &  c\, X_{1} \, \mbox{\rm sin} (X_{1})/
(L\, [1 + \frac{L s_{1}}{c}] )
\end{eqnarray}
and $X_{1}$ the nontrivial solution of $\mbox{\rm sin}(X) = \pm X/(\kappa_{0} 
\, L)$. In the second terms in (\ref{eq:fisherr2}) and
(\ref{eq:fishert2}) $s_{1}$ is 
the unstable pole, and $h_{1}$ and $h_{2}$
are the residues of the reflection and transmission amplitudes
at this pole.
After rewriting the first term in (\ref{eq:fisherr2}) and
(\ref{eq:fishert2}) as a Fourier transform,
it is straightforward to analyze ${\cal E}_{c}(0,t)$ and
${\cal E}_{p}(L,t)$ numerically. If $\kappa_{0}L \! < \! \pi /2$ 
there are no singularities with $\mbox{\rm Re}(s) > 0$, so $h_{1}=h_{2}=0$
and the second term in (\ref{eq:fisherr2}) and (\ref{eq:fishert2}) does not 
contribute. Hence, the stable regime is defined by $\kappa_{0} L < \pi/2$,
and the unstable by $\kappa_{0} L > \pi/2$. We now start by studying 
pulse reshaping by a stable PCM, and then move on to the unstable regime.

\subsection{Stable regime}
\label{sec:tempB}

\begin{figure}
\centerline{\epsfig{figure=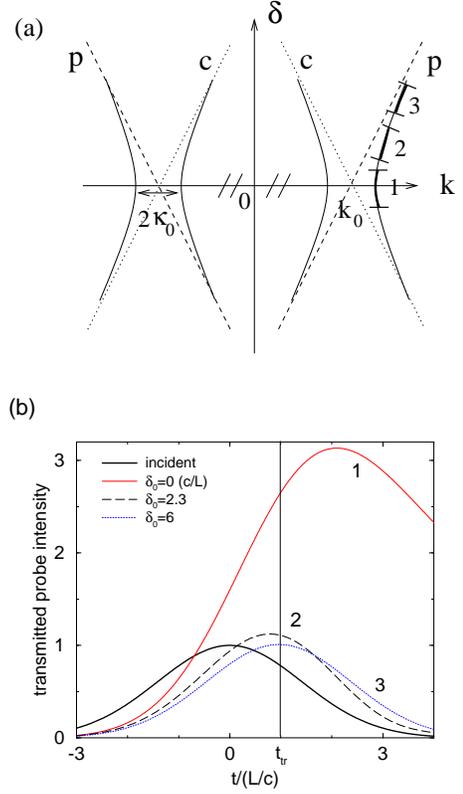,width=0.8\hsize}}
\caption[]{(a) Dispersion relation (solid line) in a PCM.
The dashed (dotted) lines correspond to the dispersion relation 
for the probe (p) and conjugate (c) waves in vacuum and
$k_{0} \equiv \omega_{0}/c$. The marked positions 1-3 indicate the 
frequency components in the incident pulses which give rise to the 
transmitted curves 1-3 in (b). (b)
Transmitted probe pulse $|{\cal E}_{p}(L,t)|$ at $x=L$ for an
incoming gaussian ${\cal E}_{p}(0,t) = e^{-\alpha t^2}
e^{i \delta_{0} t}$ at $x=0$ (thick solid line). The temporal width
(FWHM) of the incoming pulse $\Delta_{t}\equiv 2 \sqrt{\ln 2/\alpha} =
3.3\, L/c$, the spectral width $\Delta_{\delta}\equiv 4 \sqrt{\alpha \ln 2}
= 1.7\, c/L$ and $\kappa_{0} L = 1.4$.
The vertical line indicates the time $t_{tr}=L/c$ needed
to traverse the cell in vacuum.}
\label{fig:gaussstable}
\end{figure}

Fisher {\it et al.}\cite{fisher81} analyzed the phase-conjugate
reflection at a PCM in the stable regime ($\kappa_{0}L < \pi/2$). 
For a gaussian input pulse
they found a delay of the peak of ${\cal E}_{c}(0,t)$
with respect to that of ${\cal E}_{p}(0,t)$. Here we wish to emphasize
that a suitably chosen incident pulse in this stable regime 
is reshaped in such a way that
its peak emerges from the cell before the time it takes to travel the same
distance in vacuum. The central result, obtained from Eqs.~(\ref{eq:fisherr2})
and (\ref{eq:fishert2}) with $h_{1}=h_{2}=0$, is depicted 
in Fig.~\ref{fig:gaussstable}(b).
The condition on the frequencies in the input signal for 
observing the superluminal effect is that they 
should be centered away from the gap in the dispersion relation. 
The reason for this is that a pulse which is centered around 
$\delta_{0}=0$ contains
positive as well as negative frequency components whose respective positive
and negative group velocities interact and compensate in such a way that
no superluminal advancement occurs. For a pulse centered further up
on the dispersion curve (Fig.~\ref{fig:gaussstable}(a)), however, 
superluminal peak advancement is obtained
(at $\delta_{0} > 2\, c/L$, curves 2 and 3 in Fig.~\ref{fig:gaussstable}(b)). 
This superluminal peak advancement depends on three parameters:
the temporal width of the incoming pulse $\Delta_{t}$, the central
frequency of the incoming pulse $\delta_{0}$ and the coupling strength 
in the medium $\kappa_{0}$. By fixing $\kappa_{0}$ and varying
simultaneously $\Delta_{t}$ and $\delta_{0}$ the superluminal effect 
can be (numerically) optimized in several ways. In absolute terms, we find a
maximal attainable peak advancement of $~\sim 0.88\, L/c$. But since
this advancement, even though large, would only be a small
effect if the pulse is broad in time, it is also useful to optimize
the ratio $r=$ peak advancement/pulse width. We obtain a maximal 
relative peak advancement of $r\sim 0.08$.
In phase-conjugate reflection no such 
superluminal effect appears: the time at which the peak of 
the reflected signal emerges at $x=0$ is always later than $t=0$, 
the time at which the maximum of the input pulse entered the cell.

\subsection{Unstable regime}
\label{sec:tempC}

\begin{figure}
\centerline{\epsfig{figure=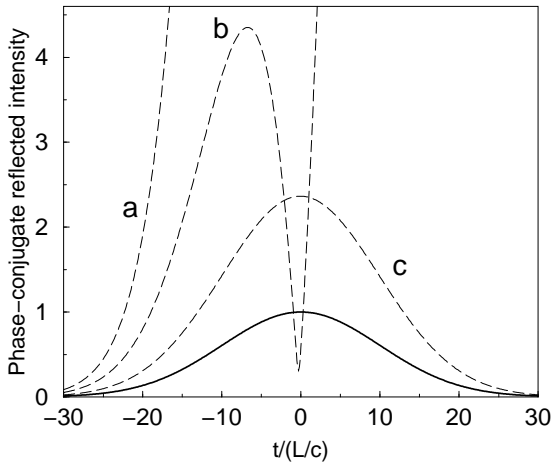,width=0.98\hsize}}
\caption[]{ 
Phase-conjugate reflected pulse $|{\cal E}_{c}(0,t)|$ at a PCM 
in its unstable operating regime ($\kappa_{0}L = 1.7$)
as a function of time (in units of $L/c$) for incident gaussian-shaped 
pulses ${\cal E}_{p}(0,t) = e^{-\alpha t^2} e^{i\delta_{0} t}$ with
different $\delta_{0}$ (in units of $c/L$). Curves a, b and c
correspond, respectively, to $\delta_{0}=0.1$, $0.28$ and $0.6$.
$|{\cal E}_{p}(0,t)|$ (thick solid line)
has temporal width $\Delta_{t}= 2\sqrt{\ln 2 / \alpha} = 24\, L/c$ and spectral 
width $\Delta_{\delta}= 4 \sqrt{\alpha \ln 2} = 0.23\, c/L$. 
}
\label{fig:gaussian}
\end{figure}

We now move on to the unstable regime ($\kappa_{0}L > \pi/2$), for
which Eqs.~(\ref{eq:fisherr2}) and (\ref{eq:fishert2}) have 
$h_{1}, h_{2} \neq 0$.
Figure~\ref{fig:gaussian} shows the 
reflected phase-conjugate pulse in this regime for an incident gaussian
centered around frequency $\delta_{0}$.

For $\delta_{0}=0.1\, c/L$ we see that the reflected signal
starts growing exponentially as soon as the incoming pulse reaches the cell.
However, for an incident pulse centered around a frequency further away
from the gap in the dispersion relation (at $\delta_{0} \approx 0.28\, 
c/L$), the reflected pulse
exhibits a local maximum before the exponential growth sets in. This peak is
clearly advanced with respect to the peak of the incoming signal.
Since $\kappa_{0} L$ is chosen close to $\pi/2$, the reflected 
pulse is greatly amplified\cite{depletion}.
 
For large $\delta_{0}$, where
the dispersion relation becomes asymptotically linear 
($\delta = \pm c (k_{0} \pm k)$), the 
reshaping of the reflected pulse is only minor and the superluminal
effect is no longer noticeable for this pulse. 
In order to optimize the superluminal response preceding exponential growth, 
one must thus have $\delta_{0}$ small (for maximum advancement) 
and $\kappa_{0} L$ close
to $\pi/2$ (for maximum intensity and delay of exponential growth).

Just as in the stable regime, we need to optimize three 
parameters simultaneously: $\kappa_{0}$, 
$\delta_{0}$ and $\Delta_{t}$. One way of doing this is by using
the exact expression (\ref{eq:fisherr2}) and analyzing it numerically, but 
this does not give much insight into the interplay between these 
parameters. Another method is to find an approximation of the exact result, 
which allows for an analytical treatment of a certain class of incident
pulses and yields quantitatively good agreement with the exact result
in that case.

In order to obtain such an approximation
we consider the conjugate reflection amplitude (\ref{eq:conjrefl}) in 
the limit $\delta \ll \kappa_{0}c$, corresponding to incident pulses 
with a large temporal bandwidth compared to $L/c$.
Equation~(\ref{eq:conjrefl}) then reduces to
\be
r_{cp}(\delta) \approx \frac{\kappa_{0}c}{\delta + i/t_{m}}
\hspace{0.6cm} \mbox{\rm for} \ \delta \ll \kappa_{0}c
\label{eq:rapprox}
\ee
with
\be
t_{m} \equiv \frac{\mbox{\rm tan} (\kappa_{0} L)}{\kappa_{0} c}.
\ee
The reflected pulse becomes in this approximation
\be
{\cal E}_{c}(0,t) \approx -i\, \kappa_{0}\, c\, e^{-\frac{t}{t_{m}}}\, \int_{-\infty}^{t} 
dt^{'} {\cal E}_{p}^{*}(0,t^{'}) e^{\frac{t^{'}}{t_{m}}}
\label{eq:conjreflapprox}
\ee
and we see that the 
growth in the unstable regime $t_{m}<0$
behaves as $e^{\frac{t}{|t_{m}|}}$. For
spectrally narrow pulses and large $\kappa_{0}$, the difference 
between (\ref{eq:conjreflapprox}) and the exact numerical result is 
found to be $< 10 \%$.

Taking the derivative of $|{\cal E}_{c}(0,t)|$ from 
Eq.~(\ref{eq:conjreflapprox}) with respect
to $t$ and equating it to zero gives a condition on the times at which
the reflected pulse is maximal ($t_{max}$) or minimal ($t_{min}$),
\be
X^2(t) + Y^2(t) - t_{m}\, \lbrack X(t) X^{'}(t) + Y(t) Y^{'}(t) \rbrack  
= 0,
\label{eq:optcond}
\ee
with 
\begin{eqnarray}
X(t) & = & \int_{-\infty}^{t} dt^{'} \mbox{\rm Re}({\cal E}_{p}^{*}(0,t^{'})) 
e^{t^{'}/t_{m}} \\
Y(t) & = & \int_{-\infty}^{t} dt^{'} \mbox{\rm Im}({\cal E}_{p}^{*}(0,t^{'})) 
e^{t^{'}/t_{m}}.
\end{eqnarray}

The optimal superluminal effect is found if the reflected intensity
is large at $t_{max}$ and close to zero at $t_{min}$ and the separation
$t_{min} - t_{max}$ is as large as possible. 
In order to obtain the maximal relative advancement
we numerically scan through the three-parameter space ($\kappa_{0}$, 
$\delta_{0}$, $\Delta_{t}$), fixing $\kappa_{0} L$ and varying the 
other two parameters
simultaneously. Using (\ref{eq:optcond}), this reveals that the 
reflected pulse is very sensitive to $\delta_{0}$ (for fixed
$\kappa_{0}$ and $\Delta_{t}$, see also Fig.~\ref{fig:gaussian}) and only 
arises for incident pulses that are sufficiently broad in time ($\Delta_{t}
\geq 20\, L/c$). For signals with broad spectra, the strong influence 
of frequency components in the instability gap prevents the formation
of a discernible pulse response before exponential growth sets in.
Furthermore, $\kappa_{0} L$ should be close to $\pi/2$, because otherwise
the fast onset of exponential growth masks the reflected pulse.
In the optimal case one can find an advancement of the peak of $t_{max} 
\sim 10\, L/c$ 
for a pulse of temporal width $\Delta_{t} \sim 25\, L/c$. The peak intensity
$|{\cal E}_{c}(0,t_{max})| \sim 5$, $|{\cal E}_{c}(0,t_{min})| \sim 0.1$
and $t_{min} - t_{max} \sim 6\, L/c$.

The results for transmission of a gaussian through an active PCM
are qualitatively the same as for phase-conjugate reflection. 
The amplitude of the advanced transmitted response is different, 
but not the values
of $\delta_{0}$ and $\Delta_{t}$ for which it arises. 
The approximation (\ref{eq:rapprox}) cannot be used
to describe the transmission in the stable regime, which occurs for 
larger values of $\delta_{0}$ than the superluminal response in the unstable 
regime. The assumption $\delta \ll \kappa_{0}\,c$ is not valid in that 
case.

\section{Pulse reflection and transmission - 
spatial analysis}
\label{sec:movie}

To gain further insight into pulse reshaping by a PCM, we follow an incoming
gaussian pulse in space. One can then observe the following stages:
(1) the probe wave packet
approaches the cell; (2) it propagates as an "optical quasiparticle"
(consisting of a probelike part traveling to the right
and a conjugatelike part traveling to the left)\cite{quasiparticles} 
in the cell and (3) 
the reflected phase-conjugate and transmitted probe packets leave the 
cell. 
We employ again the TSLT of Sec.~\ref{sec:temp}. 
The reflected phase-conjugate and transmitted probe pulses at position 
$x$ in the nonlinear 
medium and time $t$ are given by (App.~A)
\bea
{\cal E}_{c}^{\rm PCM}(x,t) & = & \frac{1}{2\pi i} 
\int_{\gamma - i\infty}^{\gamma + i\infty} ds\, h_{r} (x,is) \, 
\tilde{{\cal E}}_{p^{*}}(0,s)\, e^{st} 
\label{eq:fisherr3} \\
{\cal E}_{p}^{\rm PCM}(x,t) & = & \frac{1}{2\pi i} 
\int_{\gamma - i\infty}^{\gamma + i\infty} ds\, h_{t} (x,is) \, 
\tilde{{\cal E}}_{p}(0,s)\, e^{st},
\label{eq:fishert3}
\eea
where $h_{r}(x,\delta)$ and $h_{t}(x,\delta)$ are given by
Eqs.~(\ref{eq:conjrefl3}) and (\ref{eq:probetrans3}).
For the setup of Fig.~\ref{fig:fwm}, 
the total probe pulse ${\cal E}_{p}(x,t)$ now
consists of an incoming probe pulse in the region $x<0$, 
the probelike part of an "optical quasiparticle" in the cell ($0<x<L$) and 
a transmitted probe pulse for $x>L$. The result is 
\bea
{\cal E}_{p}(x,t) & = & \theta(-x)\ {\cal E}_{p}^{\rm PCM}(0,t-\frac{x}{c})
+ \nn \\
& & (\theta(x) - \theta(x-L))\ {\cal E}_{p}^{\rm PCM}(x,t) + \nn \\
& & \theta(x - L)\ {\cal E}_{p}^{\rm PCM}(L,t- \frac{x-L}{c}),
\label{eq:probe1}
\eea
where $\Theta$ is the Heaviside step function, and similarly,
\bea
{\cal E}_{c}(x,t) & = & \theta(-x)\ {\cal E}_{c}^{\rm PCM}(0,t+\frac{x}{c})
+ \nn \\
& & (\theta(x) - \theta(x-L))\ {\cal E}_{c}^{\rm PCM}(x,t).
\label{eq:conj1} 
\eea
We subtract from (\ref{eq:probe1}) and (\ref{eq:conj1})
again the contribution of the singularities which give rise to exponentially 
growing solutions, as in Eqs.~(\ref{eq:fisherr2}) and (\ref{eq:fishert2}), 
and rewrite them as Fourier
integrals over frequency $\delta$. We then consider an incoming gaussian
and analyze $|{\cal E}_{p}(x,t)|^2$ and 
$|{\cal E}_{c}(x,t)|^2$ as a function of x (and t) numerically, 
especially focusing on the intruiging pulse reshaping 
effects found before: the possibility for superluminal peak traversal times
in probe transmission in the stable regime and the superluminal pulse 
response in the unstably operating PCM.

\begin{figure}
\centerline{\epsfig{figure=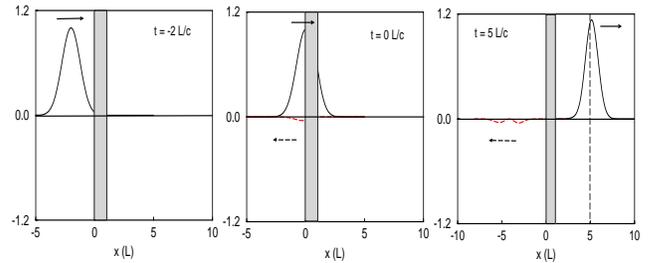,width=0.99\hsize,height=3.5cm}
\vspace{0.5cm}}
\caption{Stable regime (below-threshold) reflection and transmission of 
an incoming probe pulse at a PCM of length $L$ (indicated by shaded area). 
The solid (dashed) curve in the upper (lower) half of each time plot 
shows the probe (conjugate)
pulse $|{\cal E}_{p}(x,t)|^2$ (- $|{\cal E}_{c}(x,t)|^2$), which is
moving to the right (left). Parameters used are $\delta_{0} = 3\, c/L$,
$\Delta_{x}=2.1\, L$, $\Delta_{\delta}=1.3\, c/L$, 
$\kappa_{0} L = 1.4$. The dashed vertical line for $t=5\, L/c$ indicates 
the position where the peak of the pulse would have been in the 
absence of the PCM.
}
\label{fig:trans2}
\end{figure}

\begin{figure}
\centerline{\epsfig{figure=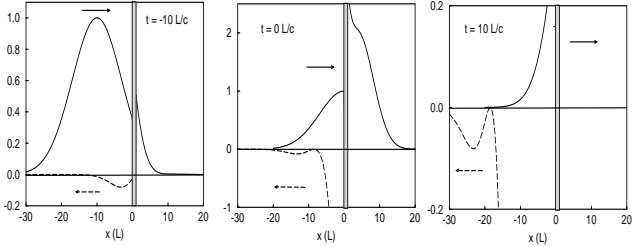,width=0.99\hsize,height=3.5cm}
\vspace{0.2cm}}
\caption{Unstable (above-threshold) regime of the PCM:
reflection and transmission of an incoming probe pulse.
The PCM is located between $0 < x < L$ (see shaded area).
The upper (lower) half of each time plot shows the probe (conjugate)
pulse $|{\cal E}_{p}(x,t)|^2$ (- $|{\cal E}_{c}(x,t)|^2$), which is
moving to the right (left). Parameters used are $\delta_{0} = 0.31\, c/L$,
$\Delta_{x}=19.3\, L$, $\Delta_{\delta}=0.144\, c/L$, 
$\kappa_{0} L = 1.7$.
}
\label{fig:trans6}
\end{figure}

Figures~\ref{fig:trans2} and \ref{fig:trans6} show the time evolution of a 
probe 
wave packet incident upon a PCM. The incoming pulse $|{\cal E}_{p}(x,t)|^2$ 
is centered around frequency $\delta_{0}$ and its width is given by 
(FWHM) $\Delta_{\delta} \equiv 2\, \sqrt{2 \alpha \ln2 }$, which 
corresponds to a spatial width $\Delta_{x} = c\, \sqrt{2 \ln2 / \alpha}$.
For clarity $-|{\cal E}_{c}(x,t)|^2$ is plotted
along the vertical axis instead of $+|{\cal E}_{c}(x,t)|^2$.

Figure~\ref{fig:trans2} depicts the probe transmission and conjugate reflection
in the stable regime ($\kappa_{0}L < \pi/2$) for an incoming pulse of spatial 
width $2.1\, L$,  
centered around frequency $\delta_{0}=3\, c/L$ in the frequency domain. 
These parameters are comparable to  
those for curve 2 in Fig.~\ref{fig:gaussstable}, but since we now consider 
$|{\cal E}_{p}(x,t)|^2$ instead of $|{\cal E}_{p}(x,t)|$, the normalization 
is different. 
We see how the probe pulse approaches the cell at $t=-2\,L/c$. At $t=0$, when
the forward tail of the pulse has entered the cell, a small reflected phase-conjugate 
pulse has developed and is traveling simultaneously in the opposite 
direction.
At $t=5\,L/c$ the advancement of the transmitted probe peak is clearly 
visible: the position of the peak is 
$x_{peak}\approx 6\, L$, whereas it would 
have been $5\, L$ if the pulse had 
propagated through vacuum (see dashed vertical line in the figure).

In the unstable regime, for $\kappa_{0} L > \pi/2$, the instability leads
to enormous growth of $|{\cal E}_{p}(x,t)|^2$ and $|{\cal E}_{c}(x,t)|^2$
as the probe pulse enters the cell. For a spectrally narrow pulse centered
around $\delta_{0}=0$, 
exponential growth
sets in immediately for that part of the probe which has reached 
the cell boundary
at $x=0$. For an incoming pulse 
centered around $\delta_{0} \sim 0.3\, c/L$ the "transient" behavior is 
recovered. 
This is illustrated in Fig.~\ref{fig:trans6}: at $t=0$
the growth of the incoming signal has set in in the PCM, but in addition 
{\it there is a clear phase-conjugate pulse with a superluminally 
advanced peak} traveling to the left. At the same time one sees
a superluminal "kink" in the transmitted probe response: the instability 
prevents the formation of a full gaussian-shaped transmitted probe pulse 
for this
set of parameters. The $t=10\,L/c$ time plot shows the superluminally reflected
signal on a larger scale.

\section{Chopped signals}
\label{sec:chop}

\begin{figure}
\centerline{\epsfig{figure=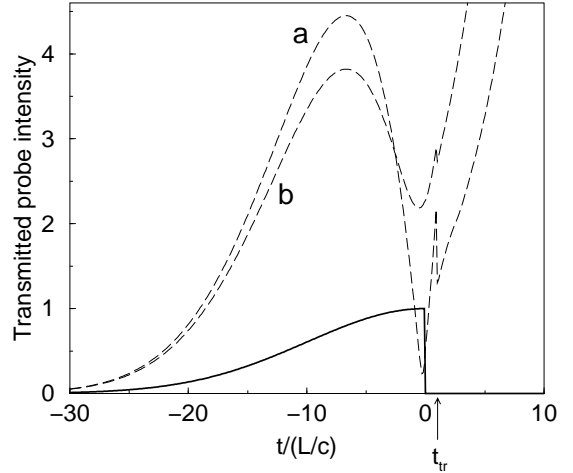,width=0.98\hsize}}
\caption{ 
Transmitted probe pulse $|{\cal E}_{p}(L,t)|$ through a PCM at $x=L$ 
for an incident gaussian-shaped
chopped pulse (thick solid line). Curves a and b correspond to
$\delta_{0}=0.28$ and $0.3$ respectively.
The parameters used are the same as in
Fig.~\protect\ref{fig:gaussian} and $t_{tr}$ indicates the time it takes
to traverse the cell in vacuum.
}
\label{fig:gausschop}
\end{figure}

Analytic gaussians are of little value as
far as information transfer is concerned, or as a check whether the 
observed superluminal effects are in agreement with 
causality\cite{japha96,bolda96,chiao94}. 
For that purpose, one needs an incoming modulated or chopped signal.
Interesting questions then arise: how is the sudden change in the input pulse 
reflected in the output pulse? How are the reflected response
and exponential growth in the unstable regime affected by this change? 
We have already shown elsewhere that for probe transmission in the 
stable regime, the edge of a chopped incoming signal is always transmitted
causally\cite{blaauboer97}.
Figure~\ref{fig:gausschop} shows 
the transmitted probe response 
in the unstable regime for an incident gaussian pulse which 
is suddenly switched off at $t=0$.
We see that the sudden change in the incoming pulse is carried over into the 
transmitted probe response at time $t_{tr} \equiv L/c$, 
the time it takes to traverse the cell in vacuum. The "information content"
of the pulse is thus transmitted with the speed of light, after which the 
exponential growth due to the instability immediately sets in.
Note that the superluminal peak advancement of the reflected conjugate pulse 
remains, just as for the full gaussian of Fig.~\ref{fig:gaussian}. This 
advancement
does not violate causality, but is a pulse reshaping effect. The fact that the 
chopped edge of the pulse is transmitted with the speed of light can be seen
more clearly by expressing ${\cal E}_{p}(x,t)$ as (App.~A)
\be
{\cal E}_{p}(L,t) = \sum_{n=0}^{\infty} \int_{-\infty}^{t - (n+\frac{1}{2}) \tau}
d t^{'} {\cal E}_{p}(0,t^{'})\, L_{n}(t, t^{'}),
\label{eq:Lfunction}
\ee
with
\be
\begin{array}{rl}
L_{n}(t, t^{'}) & = \frac{\kappa_{0} c}{2} A_{n}^{-\frac{1}{2}} 
\left(  
A_{n}^{n} I_{2n -1} \left[ \kappa_{0} c\, \sqrt{(t - t^{'})^2  
- (n+\frac{1}{2})^2 \tau^2}\, \right]  \right.
\vspace{0.2cm} \\
&  - 2 A_{n}^{n+1} I_{2n + 1}\left[ \kappa_{0} c\, \sqrt{(t - t^{'})^2 - 
(n+\frac{1}{2})^2 \tau^2}\, \right]
\vspace{0.2cm} \\
&  \left. + A_{n}^{n+2} I_{2n + 3} \left[ \kappa_{0} c\, \sqrt{(t - t^{'})^2 - 
(n+\frac{1}{2})^2 \tau^2}\, \right]
\ \ \right),
\end{array}
\ee
and
\be
A_{n} \equiv \left( \frac{t - t^{'} - (n+\frac{1}{2}) \tau }
{t - t^{'} + (n+\frac{1}{2}) \tau } \right)^{1/2}.
\ee

Here $I_{n}$ are the modified Bessel functions.
${\cal E}_{p}(L,t)$ is now expressed as a sum of integrals,
in which the $n^{th}$ term corresponds to the contribution after the $n^{th}$
round-trip time $\tau=2\, L/c$. The advantage of this expression is that it shows
that for an incoming chopped signal, which is suddenly switched on 
at $t=t_{0}$, the transmitted response only starts at $t=t_{0} + L/c$, 
in agreement with causality. 

\section{Conclusions}

In conclusion, we have theoretically studied the reflection and transmission 
of wave packets at a phase-conjugating mirror. Our main findings are

(a) In the stable operating regime 
(for $\kappa_{0} L < \pi/2$), the peak of the transmitted signal can exhibit
a superluminal peak-traversal time\cite{blaauboer97}. The conditions for
this effect are
that the incoming analytic probe signal be spectrally narrow and centered 
around a frequency sufficiently far away from the parametric resonance
($\delta=0$ in Fig.~\ref{fig:gaussstable}(a)).
The maximum advancement obtained is $\sim 0.88\, L/c$ 
and the maximum ratio (peak advancement)/(pulse width) $\sim 0.08$. 
No such effect is found in the reflected phase-conjugate response,
whose peak is always delayed with respect to the one of the incoming
probe signal (Sec.~\ref{sec:tempB}). The salient advantage of 
superluminal peak transmission
in this stable regime is that the output pulse is {\it undistorted}.

(b) In the unstable regime (for $\kappa_{0} L > \pi/2$),
for incident probe pulses with a
temporal width much larger than $L/c$, a pronounced "transient" reflected 
phase-conjugate response develops before the onset of exponential growth due 
to the instability.
This pulse response exhibits a superluminally advanced peak, both in 
phase-conjugate reflection
and in probe transmission (Sec.~\ref{sec:tempC}).
The advancement and intensity are maximal for
$\kappa_{0}L$ close to $\pi/2$ and the response is very sensitive
to the central frequency $\delta_{0}$ of the incoming pulse.

(c) We have demonstrated (Sec.~\ref{sec:chop}) that 
the superluminal features are observable only for temporally broad
analytic pulses, in agreement with the principle 
of causality, whereas a sudden (non-analytic) change in the incident 
probe pulse propagates with the speed of light.

Finally, the question arises how these pulse reshaping effects
can be observed. For a realistic PCM, consisting of a cell 
of length $L \sim 10^{-2}$~m, coupling strengths of
$\kappa_{0} c \sim c/L \sim 10^{10} s^{-1}$ have been reached,
so that tan$^2(\kappa_{0} L) \sim 1$\cite{lanzerotti96}.
In order to observe pulse reshaping in the stable operating regime 
of this PCM, one needs an incident probe pulse of width $\sim 0.1$ ns
whose peak is then transmitted with a superluminal peak transmission 
time $\sim 0.01$ ns.
To enter the unstable regime, the PCM has to be operated 
using pulsed pump beams. The pump pulses should be long enough
to allow for observation of the "transient" pulse response, but
short enough to avoid the instability effects. Since the width
of the superluminally reflected and transmitted response is on the
order of $\sim 10 \, L/c$, nanosecond pump-pulse durations 
are required.

\acknowledgements

This work was supported in part by the FOM Foundation
affiliated to the Netherlands Organization for Scientific 
Research (NWO), by the Minerva Foundation and a EU (TMR) grant.
Useful discussions are acknowledged with Iwo and Sofia Bialynicki-Birula,
R.Y. Chiao, A. Friesem and Y. Silberberg.
\vspace*{-0.5cm} \\

\appendix
\section{Two-sided Laplace transform technique}

In this appendix we briefly outline the two-sided Laplace transform 
(TSLT) technique
introduced by Fisher {\it et al.}\cite{fisher81}. The TSLT is defined as
$\tilde{\cal F}_{(*)}(x,s) \equiv \int_{-\infty}^{\infty} dt\, F^{(*)}(x,t)\, 
e^{-st}$. It is only 
valid for functions $F(x,t)$ that diminish faster than exponentially
at times $t\to -\infty$\cite{laplbook}. Its advantage compared 
to the the usual one-sided Laplace transform is that it also applies to 
functions that do not vanish at $t<0$.
The starting point of the analysis is to apply the TSLT to the 
four-wave mixing equations (\ref{eq:SEFL}) with $F(x,t) = {\cal E}_{p}(x,t)$ 
or ${\cal E}_{c}^{*}(x,t)$.
We then obtain the coupled equations
\be
\left\{ \ba{l} 
\frac{d}{dx}\, \tilde{\cal E}_{p}(x,s) + \frac{s}{c}\, \tilde{\cal E}_{p}(x,s)
+ i \kappa\, \tilde{\cal E}_{c,*}(x,s) = 0 \nonumber \vspace{0.3cm}\\
\frac{d}{dx}\, \tilde{\cal E}_{c,*}(x,s) - \frac{s}{c}\, 
\tilde{\cal E}_{c,*}(x,s)
+ i \kappa^{*}\, \tilde{\cal E}_{p}(x,s) = 0.
\ea \right.
\label{eq:LaplSEFL}
\ee
Equations (\ref{eq:LaplSEFL}) are solved together with the Laplace 
transforms of 
the boundary conditions ${\cal E}_{p}(0,t) = F(0,t)$, where
$F(0,t)$ is an incident probe pulse
at the entry $x=0$ of the PCM medium,  
and ${\cal E}_{c}(L,t) = 0$, so no incoming conjugate pulse at the end 
of the medium. The result is
\bea
\tilde{\cal E}_{c}(x,s) & = & h_{r}(x,is)\, \tilde{\cal F}_{*}(0,s)
\nonumber \\
\tilde{\cal E}_{p}(x,s) & = & h_{t}(x,is)\, \tilde{\cal F}(0,s),
\nonumber
\eea
with the reflection and transmission amplitudes
\bea
h_{r} (x,\delta) & = & \frac{\kappa_{0} \sin(\beta (L-x))} {\frac{\delta}{c}
\mbox{\rm sin} (\beta L) + i \beta \mbox{\rm cos} (\beta L)}
\label{eq:conjrefl3} \vspace{0.3cm} \\
h_{t} (x,\delta) & =  & \frac{i \beta \cos (\beta (L-x)) + \frac{\delta}{c}
\sin (\beta (L-x))}
{\frac{\delta}{c}\mbox{\rm sin} (\beta L) + i \beta \mbox{\rm cos} 
(\beta L)}, 
\label{eq:probetrans3}
\eea
where $\beta$ is given by Eq.~(\ref{eq:betavec}).

The reflected phase-conjugate pulse 
${\cal E}_{c}(x,t)$
and transmitted probe pulse ${\cal E}_{p}(x,t)$ at position $x$
in the nonlinear medium and time $t$ are then obtained
by using the inverse Laplace transform. At $x=0$ and $x=L$, respectively, 
they are given by 
\bea
{\cal E}_{c}(0,t) & = & \frac{1}{2\pi i} \int_{\gamma - i\infty}^{\gamma + 
i\infty} ds\, r_{cp} (is) \, \tilde{{\cal E}}_{p^{*}}(0,s)\, e^{st} 
\label{eq:fisherr} \\
{\cal E}_{p}(L,t) & = & \frac{1}{2\pi i} \int_{\gamma - i\infty}^{\gamma + 
i\infty} ds\, t_{pp} (is) \, \tilde{{\cal E}}_{p}(0,s)\, e^{st},
\label{eq:fishert}
\eea
where $r_{cp} = h_{r}(x=0)$ and $t_{pp} = h_{t}(x=L)$.
The choice of contour $\gamma$ in (\ref{eq:fisherr}) and
(\ref{eq:fishert}) is in agreement with causality, which means in this 
case that it is to the right of all 
the singularities of $r_{cp}(is)$ (and $t_{pp}(is)$).
The singularities in the right-half s-plane give rise to exponential growth of
${\cal E}_{c}(x,t)$ and ${\cal E}_{p}(x,t)$. The integrals (\ref{eq:fisherr})
and (\ref{eq:fishert}) can be evaluated by taking the contribution of these
poles separately and rewriting the remaining integral as a Fourier
integral, see (\ref{eq:fisherr2}) and (\ref{eq:fishert2}).

They can also be evaluated in another way, which is especially
insightful when regarding non-analytic, chopped probe pulses
(Sec.~\ref{sec:chop}). To that end we rewrite
(\ref{eq:fisherr}) and (\ref{eq:fishert}) as
\bea
{\cal E}_{c}(0,t) & = & \int d t^{'}\, {\cal E}_{p}^{*}(0, t^{'}) 
H_{c}(t, t^{'}) \\
{\cal E}_{p}(L,t) & = & \int d t^{'}\, {\cal E}_{p}(0, t^{'}) H_{p}(t, t^{'}),
\label{eq:probeana}
\eea
with
\bea
H_{c}(t, t^{'}) & = & \frac{1}{2\pi i}\int_{\gamma - i\infty}^{\gamma + i\infty}
ds\, r_{cp}(is)\, e^{-s(t^{'} - t)} \\
H_{p}(t, t^{'}) & = & \frac{1}{2\pi i}\int_{\gamma - i\infty}^{\gamma + i\infty}
ds\, t_{pp}(is)\, e^{-s(t^{'} - t)}
\label{eq:Hana}
\eea
In order to evaluate the integral in, e.g., $H_{p}(t, t^{'})$, $t_{pp}$ is 
first rewritten as the series
\be
t_{pp}(is) = 2\eta \sum_{n=0}^{\infty} \frac{a^{2n}}{(s+\eta)^{2n+1}}\,
e^{-(n + \frac{1}{2})\eta \tau}
\label{eq:series}
\ee
with $a\equiv \kappa_{0} c$, $\eta \equiv \sqrt{s^2 - a^2}$ and $\tau = 2L/c$, 
the roundtrip time of the PCM. One can then easily prove that (\ref{eq:series})
is uniformly convergent, which allows for term by term integration of 
(\ref{eq:Hana}). We use the substitution
\be
s = \frac{i a}{2} \left( \frac{u}{A_{n}} - \frac{A_{n}}{u} \right)
\ee
with
\be
A_{n} \equiv \left( \frac{t - t^{'} - (n+\frac{1}{2}) \tau }
{t - t^{'} + (n+\frac{1}{2}) \tau } \right)^{1/2},
\ee
in which $t^{'} + (n+\frac{1}{2})\tau$ is the retardation time after
$(n+\frac{1}{2})$ round trips. We then arrive at the result
Eq.~(\ref{eq:Lfunction}).

Similarly, the conjugate reflected pulse is given by
\bea
{\cal E}_{c}(L,t) & = & \int_{-\infty}^{t} d t^{'}\, 
{\cal E}_{p}^{*}(0,t^{'})\, M_{0}(t, t^{'}) + \nonumber \\
& & - 2 \sum_{n=1}^{\infty} \int_{-\infty}^{t - n \tau}
d t^{'} {\cal E}_{p}^{*}(0,t^{'})\, M_{n}(t, t^{'}),
\label{eq:Lfunctionconj}
\eea
with
\be
\begin{array}{rl}
M_{n}(t, t^{'}) & = \frac{-i\, \kappa_{0} c}{4}  
\left(  
B_{n}^{n-1} I_{2n -2} \left[ \kappa_{0} c\, \sqrt{(t - t^{'})^2  
- n^2 \tau^2}\, \right]  \right.
\vspace{0.2cm} \\
&  - 2 B_{n}^{n} I_{2n}\left[ \kappa_{0} c\, \sqrt{(t - t^{'})^2 - 
n^2 \tau^2}\, \right]
\vspace{0.2cm} \\
&  \left. + B_{n}^{n+1} I_{2n + 2} \left[ \kappa_{0} c\, \sqrt{(t - t^{'})^2 - 
n^2 \tau^2}\, \right]
\ \ \right),
\end{array}
\ee
and
\be
B_{n} \equiv \left( \frac{t - t^{'} - n\, \tau }
{t - t^{'} + n\, \tau } \right).
\ee


\begin{references}
\bibitem{depletion} If the intensities of probe and conjugate waves approach
		those of the two pumps, pump depletion has to be
		taken into account and our analysis is not valid anymore.
\end{references}
\end{document}